\documentclass[letterpaper,twocolumn,10pt]{article}
\usepackage{stylefile}
\usepackage{listings}
\usepackage{color}
\usepackage{xspace}

\usepackage{tikz}
\usepackage{amsmath}
\usepackage{multirow}

\newcommand{\eg}{\textit{e.g.,}\xspace}
\newcommand{\ie}{\textit{i.e.,}\xspace}

\newcommand{\myparagraph}[1]{\vspace{1mm} \smallskip \noindent{\bf {#1}}}

\definecolor{WhiteBlock}{RGB}{250,250,250}
\definecolor{BlackText}{RGB}{0,0,0}
\definecolor{BlueTypename}{RGB}{0,0,204}
\definecolor{GreenString}{RGB}{96,172,57}
\definecolor{PurpleKeyword}{RGB}{204,0,204}
\definecolor{GrayComment}{RGB}{140,140,140}
\definecolor{GoldDocumentation}{RGB}{180,165,45}
\lstdefinelanguage{sail}
{
    columns=fullflexible,
    keepspaces=true,
    frame=single,
    framesep=0pt,
    framerule=0pt,
    framexleftmargin=1pt,
    framexrightmargin=1pt,
    framextopmargin=2pt,
    framexbottommargin=2pt,
    xleftmargin=12pt,
    xrightmargin=3pt,
    numbers=left,
    numberstyle=\small\color{GreenString},
    numbersep=3pt,
    backgroundcolor=\color{WhiteBlock},
    basicstyle=\ttfamily\color{BlackText},
    keywords={
    type,val,let,in,function,scattered,enum,union,clause,default,order,dec,register,vector,bitfield
    },
    keywordstyle=\color{PurpleKeyword},
    ndkeywords={
        if,then,while,for,return,match,else,
        atom,int,forall,effect,infix,overload,operator,
        =>,>=,<=,=,->,-,:,,,+,*,.,@
    },
    ndkeywordstyle=\color{BlueTypename},
    stringstyle=\color{GreenString},
    string=[b]"
}

\lstdefinelanguage{smtlib}
{
    columns=fullflexible,
    keepspaces=true,
    frame=single,
    framesep=0pt,
    framerule=0pt,
    framexleftmargin=1pt,
    framexrightmargin=1pt,
    framextopmargin=2pt,
    framexbottommargin=2pt,
    xleftmargin=12pt,
    xrightmargin=3pt,
    numbers=left,
    numberstyle=\small\color{GreenString},
    numbersep=3pt,
    backgroundcolor=\color{WhiteBlock},
    basicstyle=\small\color{BlackText},
    keywords={
    trace, define, enum, assume, reg, read, write
    },
    keywordstyle=\color{PurpleKeyword},
    ndkeywords={
        nil, field, bits, struct
    },
    ndkeywordstyle=\color{BlueTypename},
    stringstyle=\color{GreenString},
    string=[b]"
}

\lstdefinelanguage{sailor-output}
{
    columns=fullflexible,
    keepspaces=true,
    frame=single,
    framesep=0pt,
    framerule=0pt,
    framexleftmargin=1pt,
    framexrightmargin=1pt,
    framextopmargin=2pt,
    framexbottommargin=2pt,
    xleftmargin=12pt,
    xrightmargin=3pt,
    numbers=left,
    numberstyle=\small\color{GreenString},
    numbersep=3pt,
    backgroundcolor=\color{WhiteBlock},
    basicstyle=\ttfamily\color{BlackText},
    keywords={
    Read, Write, R, W, false, true
    },
    keywordstyle=\color{PurpleKeyword},
    ndkeywords={
        ISA, state, footprint, user, executable, supervisor, machine, 
    },
    ndkeywordstyle=\color{BlueTypename},
    morecomment=[s][\color{GrayComment}\slshape]{(}{)},
    stringstyle=\color{GreenString},
    string=[b]"
}

\begin{document}

\date{}

\title{\Large \bf Automatic ISA analysis for Secure Context Switching}

\author{
    \begin{tabular}{c c}
        {\rm Neelu S.\ Kalani \thanks{Majority of the work was done while the author was at IBM Research --- Zurich.}} & {\rm Thomas Bourgeat} \\
        EPFL, Switzerland \& IBM Research --- Zurich & EPFL, Switzerland \\
        & \\ %
        {\rm Guerney D.\ H.\ Hunt} & {\rm Wojciech Ozga} \\
        IBM T. J. Watson Research Center & IBM Research --- Zurich
    \end{tabular}
} %

\maketitle

\begin{abstract}

Instruction set architectures are complex, with hundreds of registers and instructions that can modify dozens of them during execution, variably on each instance. Prose-style ISA specifications struggle to capture these intricacies of the ISAs, where often the important details about a single register are spread out across hundreds of pages of documentation. 
Ensuring that all ISA-state is swapped in context switch implementations of privileged software requires meticulous examination of these pages.
This manual process is tedious and error-prone.  

We propose a tool called Sailor that leverages machine-readable ISA specifications written in Sail to automate this task. 
Sailor determines the ISA-state necessary to swap during the context switch using the data collected from Sail and a novel algorithm to classify ISA-state as security-sensitive. 
Using Sailor's output, we identify three different classes of mishandled ISA-state across four open-source confidential computing systems. 
We further reveal five distinct security vulnerabilities that can be exploited using the mishandled ISA-state. 
This research exposes an often overlooked attack surface that stems from mishandled ISA-state, enabling unprivileged adversaries to exploit system vulnerabilities.

\end{abstract}

\section{Introduction}
\label{sec:intro}
The increasing complexity of architectures, and the nitty-gritty details of the Instruction set architectures (ISA) being scattered throughout hundreds or thousands of pages have put system security at risk. 
A simple overlook of a single register in security-critical code can inadvertently expose the system to security vulnerabilities. 
Low-level software developers need to do the tedious job of manually navigating through the dense ISA specifications to ensure they do not introduce any security vulnerabilities in their code. 
Moreover, developers need to stay up-to-date with the evolving ISA specifications and the hardware platforms that support them.

ISAs have traditionally been specified using hand-written semantics in natural language, occasionally accompanied by pseudo-code, in bulky and comprehensive documents. 
For example, the Arm A-profile architecture reference manual~\cite{arm-ref-manual} and the privileged and unprivileged RISC-V ISA manuals~\cite{riscv-isa} consist of $~$15,000 and $~$900 pages, respectively. 
These ISA manuals are highly technical and are becoming increasingly dense, with new features and instructions added in every version. 
For instance, the Arm v7-A, v8-A and v9-A consist of $~$2700, $~$5200, and $~$15,000 pages, respectively~\cite{arm-v7-a, arm-v8-a, arm-ref-manual}. 
Similarly, on RISC-V, the non-ISA specifications (interrupt architecture, IOMMU, etc.) further add 767 pages\cite{riscv-specs}, and new ISA extensions are ratified every year~\cite{riscv-ratified}. 
Prior work shows a similar trend for the x86 ISA~\cite{hotos-baumann}. 

\begin{figure}
    \centering
    \includegraphics[width=160pt]{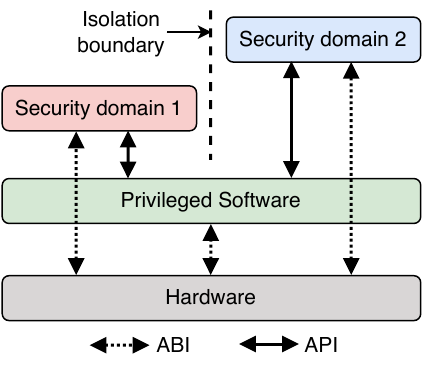}
    \caption{Security domains isolated by a privileged software that mediates the context switch between the domains. 
    The security domains access the ISA-state through the Application Binary Interface (ABI). The privileged software must swap all security-sensitive ISA-state to prevent attacks.} 
    \label{fig:conf-comp}
\end{figure}

New features introduce new ISA-state, adding more complexity to the already convoluted ISA-state, especially when it comes to isolating software. \autoref{fig:conf-comp} shows two different security domains that execute in a system.
A privileged software mediates the context switch between the two security domains.
To the security domains, it should appear as if they are the only software that executes in the system~\cite{enclave-isolation, certikos}. 
There are two layers at which this guarantee needs to be enforced: the application programming interface (API) and the application binary interface (ABI). 
At the API-layer, which enables interaction between the different security domains, inputs from untrusted sources requires sanitization to prevent Iago-style attacks \cite{iago}. 
Through the ABI-layer, the security domains can access a significant part of the ISA-state. 
This involves 1000+ and 140+ registers on Arm v9-A~\cite{sail-arm} and RISC-V~\cite{sail-riscv}, respectively. 
The privileged software must ensure that all security-sensitive data of the first security domain has been wiped clean from the ISA-state, before passing control to the second security domain.

In this paper, we focus on the ABI-layer, improper sanitization at which can introduce various security vulnerabilities. 
Prior work has discovered security vulnerabilities at the ABI-layer in industry-standard software as well as academic research projects. 
These involve improper flags sanitization \cite{totw} in Intel SGX-SDK~\cite{sgx-sdk} and Microsoft's open enclave~\cite{ms-open-enclave} runtimes,
register leakage~\cite{guard-dilemma} and ROP~\cite{rop-sgx} on Intel SGX-LKL~\cite{sgx-lkl} (another SGX runtime), interrupt hijacking~\cite{intel-tdx-sec} on Intel TDX~\cite{intel-tdx}, side-channels and computational integrity attacks through the floating point unit on Intel SGX and RISC-V Keystone \cite{dtrap-fpu, keystone}. 
Small oversights (of even a single bitfield in a register) in the context switch implementation introduce high-impact vulnerabilities that lead to simple and cost-effective attacks that even unprivileged adversaries can mount. 
The matter gets worse when dealing with ISAs such as RISC-V that promote adding new extensions and the privileged software inadequately tracks which ISA extensions are enabled by the platform vs. supported by the software \cite{dtrap-fpu}.

Essentially, the attack surface introduced by improper sanitization of ISA-state (specifically, general purpose registers (GPRs) and control and status registers (CSRs)) when switching across security domains is often overlooked. 
To approach this problem, we propose leveraging the machine-readable ISA specification written in the Sail language~\cite{sail} to capture ISA-level insights and compute the set of security-sensitive ISA-state. 
With that, we aim to equip privileged software developers with conveniently accessible insights to the ISA that eliminates the need for developers to be overwhelmed with hundreds or thousands of pages of manuals. 

Sail enables specifying the semantics of all instructions in the ISA, capturing the effects of executing instructions in a complex system with multiple privilege levels on the ISA-state. 
Sail has already been used for writing specifications of multiple architectures (x86, Arm, RISC-V, MIPS, CHERI, and IBM Power).   
Sail has also resulted in tools such as Isla \cite{isla} for symbolic execution of Sail code using the Z3 SMT Solver, and verification of machine code using Islaris \cite{islaris}.

We introduce Sailor, a novel tool that computes the set of security-sensitive ISA-state by automatically analyzing the ISA specification in Sail~\cite{sail}.
Sailor eliminates the tedious task of navigating through complex ISA specifications by leveraging Sail models of ISAs
to automatically compute the set of ISA-state that should be swapped during context switches across security domains (security-sensitive ISA-state). 
We apply Sailor on the Sail model for the RISC-V architecture to identify security-sensitive ISA-state \cite{sail-riscv}. 
Sailor parses the Sail model to automatically extract ISA-level insights that reflect the effects of executing instructions on the ISA-state as well as the effects of the ISA-state on instruction execution. 
For example, exceptions that occur during floating-point instruction execution on RISC-V update the \texttt{fflags} as a side-effect. 
Prior work demonstrated a side-channel attack where an adversary infers the victim's floating-point operations and exceptions by observing the \texttt{fflags} CSR~\cite{dtrap-fpu}. 
While prior work relied on manual ISA inspection, Sailor automates the discovery of such ISA-level insights.
We leverage these insights to compute the set of security-sensitive ISA-state and use that to perform security analysis of \textbf{four open-source confidential computing systems} i.e. Keystone \cite{keystone}, Komodo~\cite{komodo}, IBM ACE \cite{ibm-ace}, and Rivos Salus \cite{rivos-salus}. 
We find \textbf{three different classes of mishandled ISA-state} in the context switch implementations
of these systems.
We further reveal five different types of vulnerabilities that can be exploited across the systems by leveraging the mishandled ISA-state: computational integrity attacks, misconfigured emulated I/O fences, side-channels~\footnote{Side-channels in the context of this paper refers to exploiting ISA-state to unintentionally leak information. It does not refer to micro-architectural side-channels.}, covert-channels, and timing-channels (\autoref{tab:vuln}).~\footnote{Prior work previously reported two of these vulnerabilities for the Keystone framework~\cite{dtrap-fpu}.}

Our main contributions are: 
\begin{enumerate}
    \item We design and implement a new tool called Sailor that performs automatic analysis of the ISA specification written in Sail to compute the set of ISA-state that must be swapped during context switches.
    \item We evaluate four open-source confidential computing frameworks using the results of the algorithm. We report three different classes of mishandled ISA-state %
    in the context switch code of the frameworks (\autoref{tab:vuln}). We identify five different types of security vulnerabilities that arise from exploiting these mishandled ISA-state.  
    \item We validate the ISA-level insights generated from Sailor using an existing symbolic execution tool for Sail, Isla \cite{isla}. We found one bug in our tool during this process. 
\end{enumerate}

\textbf{Responsible disclosure:}  
We have disclosed all of the bugs and insights discovered through this work to each of the open-source frameworks (which are all research projects and are not used in production) via appropriate communication channels.

\section{Motivation \& Background}
\label{sec:motivation}
\subsection{ISAs are Complex}

ISA manuals are increasingly dense and ISA-state (architectural context), increasingly convoluted. 
It primarily comprises of general purpose registers (GPRs), control and status registers (CSRs), and memory. 
There is a plethora of research on keeping memory of different security domains confidential using various access control mechanisms \cite{vtx, riscv-isa, arm-mpu, sgx, trustzone}.
Yet, little focus has been put towards preventing security vulnerabilities that arise from improper sanitization of ISA-state when switching across security domains~\cite{totw}.  
However, high-impact vulnerabilities can be trivially introduced in context switch implementations while swapping ISA-state, that can further be exploited by an unprivileged adversary with minimal effort. 
The reason for this is two-fold.

First, control and status registers (CSRs in RISC-V), special-purpose registers (Arm) or model specific registers (MSRs in x86) are immensely complex - up to the bit-level. 
Each bit in such registers can reconfigure the processor's behavior, thus influencing the execution of the following instructions.
It is trivial to introduce bugs when setting/clearing some of the bits in a register due to possible inter-dependencies with bits of another register. 
For instance, when any floating point (FP) operations are executed on RISC-V platforms, a Dirty bit is set in the \texttt{sstatus} register as an optimization to check whether the floating point register state needs be saved. 
However, not clearing this bit during a context switch can leak information about floating point unit use inside a security domain. 
Similarly, prior work reported that Intel SGX's SDK fails to clear two bits during the context switch that harm the computational integrity of the operations executed inside the enclave and also lead to memory corruption~\cite{totw}. 

Second, there is a higher risk of introducing vulnerabilities at the ISA-level for ISAs that promote modularity and customization, such as RISC-V.
This is especially true if the privileged software does not properly handle discovery of the ISA extensions supported by the platform on which it runs. 
Prior work shows that incorrect sanitization of floating point state on the x86-64 and RISC-V architectures leads to unintentional side-channels that can be exploited by unprivileged adversaries~\cite{dtrap-fpu}. 
Moreover, one of the three classes of mishandled ISA-state we find also stem from ISA-extensions. 

\subsection{Overlooking security at the ABI-layer}

Existing strategies to avoid vulnerabilities focus on preventing software bugs at the application level (e.g. application binary scanning, symbolic execution of user code, etc.) or introduce patches to prevent sophisticated micro-architectural attacks (speculative execution, cache side-channels)~\cite{spectre, meltdown,flush-reload}. 
These techniques do not target bugs at the ABI-layer that can breach the security guarantees of security domains even when the application code is secure and caches are partitioned. 

Further, formal verification of privileged software (that targets to avoid attacks at the ABI-layer too) relies on self-written specifications (rather than official ISA specifications) that usually model a subset of the entire machine \cite{serval, komodo, certikos}.
This is important to make the verification efforts feasible, by limiting and pruning the paths that need to be explored. 
These partial machine models are used to prove functional correctness of the security monitors \ie{} whether the context switch is saving and restoring the registers (to and from memory) as expected.
They are also used to prove the non-interference security property \ie{} on switches from security domain A to security domain B and then back to security domain A, the security domain A must not be able to distinguish the system state before vs. after the execution of security domain B. 
However, the verification only proves properties that hold for the ISA-state modeled by the self-written specifications, instead of the entire ISA-state. 
In such cases, it's difficult to extend the proven properties to the ISA-state that was not modeled.
It is also difficult to identify security-sensitive ISA-state in the first place, which could be useful to determine the exact subset of the ISA that the verification frameworks must model. 
However, that requires studying the ISA specifications extensively and meticulously. 

\subsection{Secure Context Switches}
A naive approach to implement secure context switches is to consider all ISA-state to be intrinsically security-sensitive.
Thus, a context switch implementation in privileged software would swap the entire ISA-state when switching across security domains. 
While in theory this approach seems intuitive and appealing, there are two problems with this approach.

First, due to the convoluted nature of architectures, with hundreds or thousands of registers~\cite{riscv-isa, arm-ref-manual}, it is trivial to overlook ISA-state that requires to be swapped during context switches across security domains. 
This is especially true in the case of ISA-extensions. 
It is because of such oversights that security vulnerabilities get a chance to surface-up and be exploited by unprivileged adversaries with minimal efforts~\cite{dtrap-fpu, totw}. 

Second, context switches are expensive~\cite{context-switch-cost, context-switch-cost-hotos}.
Context switching across security domains doesn't just involve reading and writing of registers, but also saving the current context onto memory and reading the previously stored context from memory. 
Further, depending on the ISA-state that is exposed to a security domain (executing in a particular privilege mode) and the ISA-state that the execution of the security domain is dependent on, not all ISA-state might be security-sensitive. 
For example, there are three privilege modes on RISC-V \ie{} User (U-mode), Supervisor (S-mode) and Machine (M-mode).
If two security domains execute in U-mode, then the S-mode CSR \texttt{satp} would be considered security-sensitive as it contains the page table root address, that would impact the execution of the security domains in U-mode. 
However, the S-mode CSR \texttt{sepc} that contains the exception program counter on traps from U-mode to S-mode, would not be considered security-sensitive since it neither impacts execution of the security domains in U-mode, nor is accessible in U-mode.  
It can be beneficial especially for performance-critical systems to dig deeper in the ISA and find out exactly which ISA-state needs to be swapped during a context switch.

\section{Sailor Design}
\label{sec:sailor} 

\begin{figure}
    \includegraphics[width=\columnwidth]{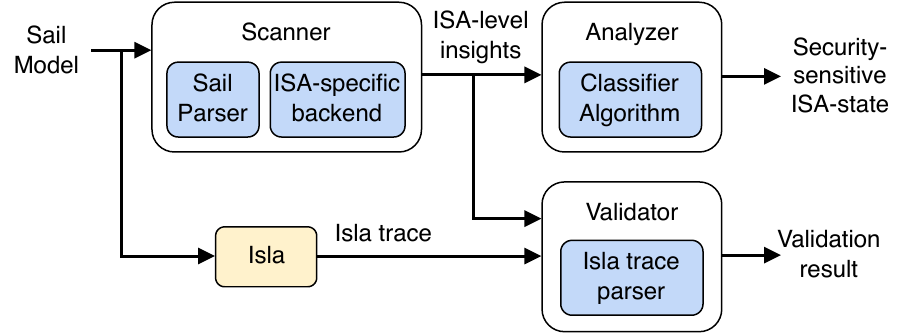}
    \caption{Sailor Overview}
    \label{fig:sailor}
\end{figure}

\autoref{fig:sailor} shows the high-level overview of Sailor. It consists of three components: (A) Scanner, (B) Analyzer, and (C) Validator. The ultimate goal of the tool is to enable secure context switching of ISA-state by software that intends to enforce isolation between different security domains running on the system. 

The Scanner comprises of our Sail parser and an ISA-specific backend. 
The Scanner parses the Sail model of the ISA. 
This involves discovering the ISA, \ie{} privilege modes, registers and instructions. 
Then the Scanner dives deeper into the Sail model by parsing the register and instruction definitions, to extract their semantics and the inter-dependencies between different parts of the ISA.

The Analyzer takes as input the ISA-level insights extracted by the Scanner. 
The \texttt{Classifier Algorithm} (\autoref{sec:analyzer}) processes this input to classify ISA-state as security-sensitive, that must be swapped during context switches.

The quickest approach to leverage the algorithm's output is to use utilities like grep to find out whether existing software swaps the security-sensitive ISA-state. 
In this scenario, Sailor serves as a compass to bring the developer's attention to where the code manages the security-sensitive ISA-state, if it exists, and check that it does so properly.
Software developers can also use it as a guide for writing privileged software. 
For ISAs like Arm that could have hundreds of CSRs in the security-sensitive ISA-state, this information can be used to automatically generate tests. 
The tests can be simple assembly code that checks the security-sensitive ISA-state and rereads it back after a context switch to check if it changed. 
The algorithm's output can also be leveraged by ISA architects that design new ISA extensions as a guideline to ensure that the proposed extension does not affect the previously determined security-sensitivity of the ISA-state. We discuss this further in Section \ref{sec:isa-ext-discussion}.

Finally, the Validator corroborates the output of the Scanner using an existing tool for symbolic execution of Sail code, Isla~\cite{isla}.

\subsection{Scanner} 

The Scanner automatically extracts insights from the ISA specifications written in machine-readable Sail code. 

\subsubsection{Primer on Sail}

\begin{figure}
    \centering
    \includegraphics[width=\columnwidth]{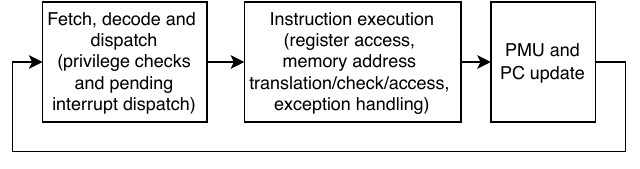}
    \caption{RISC-V Sail Model implementation flow block diagram}
    \label{fig:sail-riscv}
\end{figure}

\autoref{fig:sail-riscv} shows the block diagram for the instruction fetch till execute loop in the RISC-V Sail implementation \cite{sail-riscv}. 
Each instruction is first fetched, decoded and then dispatched. Before the instruction semantics are executed, two things are checked: 
(i) if any interrupts are pending, which might be serviced before executing the instruction, and 
(ii) if the privilege level in the current architectural context allows executing the instruction.

Each instruction (or sometimes group of the same type of instructions, depending on the Sail model implementation) has an independent definition specified in the Sail model implementation. 
This definition of the instruction is then executed, which may involve accessing memory (and corresponding memory translations and checks), performing arithmetic operations, or modifying the processor state (registers). 
The execution may also lead to exceptions, in which case the appropriate handler must be called. 
Once the execution is complete, the performance monitoring unit (PMU) and the PC are updated, and the procedure is repeated for the next instruction. 

We will now deep dive into a small Sail snippet to understand how instruction specifications are implemented in Sail. 
The following text is taken from the RISC-V privileged ISA specification manual \cite{riscv-isa}, which describes the \texttt{mret} instruction.  

\textit{"An MRET or SRET instruction is used to return from a trap in M-mode or S-mode respectively. 
When executing an xRET instruction, supposing xPP holds the value y, xIE is set to xPIE; the privilege mode is
changed to y; xPIE is set to 1; and xPP is set to the least-privileged supported mode (U if U-mode is
implemented, else M). 
If y$\neq{}$M, xRET also sets MPRV=0."} 

\begin{lstlisting}[language=sail,caption={Sail implementation of \texttt{mret} in RISC-V.},captionpos=b,label={lst:mret}]
function clause execute MRET() = {
  if   cur_privilege != Machine
  then { handle_illegal(); RETIRE_FAIL }
  else {
    mstatus[MIE]  = mstatus[MPIE];
    mstatus[MPIE] = 0b1;
    cur_privilege = privLevel(mstatus[MPP]);
    let prev_priv = if extEnabled(Ext_U)
                    then User else Machine);
    mstatus[MPP]  = bits(prev_priv);
    if   cur_privilege != Machine
    then mstatus[MPRV] = 0b0;
    PC = prepare_xret_target(Machine)
    RETIRE_SUCCESS
  }
}
\end{lstlisting}

\autoref{lst:mret} shows the corresponding Sail implementation of the \texttt{mret} instruction. 
The instruction definition is implemented using the "function clause execute <name>()", which we see on the left side of the figure. 
The instruction execution is allowed only if the privilege mode in the current architecture context is set to machine mode (M-mode). 
The exception handler is then called to execute the "CTL\_MRET" case (depicted on the right side of the figure). 
Here, we can see the translation of the prose-style specification listed above into machine-readable specification in Sail.
For instance, \texttt{MIE} (machine interrupt enable) is set to \texttt{MPIE} (machine previous interrupt enable) and the privilege level is set to the value in \texttt{MPP} (Machine previous privilege level) in the \texttt{mstatus} register. 

The Sail implementation thus describes, in machine-readable format, every single effect on the architectural context of executing an instruction, given the access rights of a particular privilege mode.

\subsubsection{Capturing ISA-level insights}
\label{sec:scanner-insights}
We formulate specific questions that help extract ISA-level insights, specifically about the access rights of different privilege modes over the ISA-state and the side-effects of executing instructions on the ISA-state. 
The Scanner answers the following questions about the ISA. 

\textbf{1. Does a privilege mode have explicit access rights to read or write a particular ISA-state?} 
The simplest method to establish a side-channel or a covert-channel is through shared ISA-state. 
For example, on RISC-V, confidential VMs that execute in VS-mode have the permission to explicitly write the \texttt{vsscratch} CSR; and
hypervisors that execute in HS-mode have the permission to explicitly read the \texttt{vsscratch} CSR. 
Thus, if a security monitor does not clear the \texttt{vsscratch} CSR during a context switch, it can be leveraged to leak the confidential VM's security-sensitive information. 
In other words, this question checks whether a privilege mode can execute explicit ISA-state read/write instructions (\eg{} \texttt{csrrs} and \texttt{csrrw} in RISC-V from the Zicsr extension).
The answer to this question gives us insights that the Analyzer uses to determine whether a direct communication channel can be established between two different security domains. 

\textbf{2. What is the ISA-state footprint of an instruction?} 
We define the ISA-state that could be read or written as a side-effect of the execution of an instruction as the \textit{ISA-state footprint} of an instruction. 
For example, as shown in \autoref{lst:mret}, executing the \texttt{mret} instruction has several side-effects on the ISA-state, \eg{} updating the \texttt{MIE} field with the \texttt{MPIE} field's value in the \texttt{mstatus} CSR. 
This implies that, the \texttt{mret} instruction implicitly depends on the value of \texttt{MPIE} and updates the value of \texttt{MIE}. 
This information is crucial as instructions implicitly (or indirectly) update ISA-state.
When the context switch implementation does not clear such ISA-state, a security domain can infer information about the execution in another security domain. 
For example an adversary can infer whether the victim uses the floating point unit and whether any (and which) exceptions occur during execution~\cite{dtrap-fpu}.
We also include the interrupt handling, PMU and PC updates in the footprint (as shown in Figure \ref{fig:sail-riscv}). 
Prior work has shown that security vulnerabilities like interrupt hijacking can stem from overlooking a even a few bits of a register in the context switch implementation\cite{intel-tdx-sec}. 
Thus, it is crucial to consider security-sensitive ISA-state that changes as a side-effect of instruction execution during the context switch implementation. 

\textbf{3. Does a privilege mode have the access rights to execute a particular instruction?} 
If a privilege mode does not have the access rights to execute an instruction (\eg{} only M-mode can execute the \texttt{mret} instruction) and ultimately never indirectly reads or writes certain ISA-state, then that helps reduce the amount of ISA-state that needs to be swapped to perform a secure context switch.
We already cover direct reads or writes of ISA state in the first question.
This information can be critical, especially performance-wise, as context switches are known to be expensive~\cite{context-switch-cost, context-switch-cost-hotos}.

\autoref{lst:mret-sailor} shows the ISA-level insights that Sailor generates for the \texttt{mret} instruction. 
This includes per privilege mode access to 
executing the instruction as well as a fine-grained ISA-state footprint. 
The ISA-state footprint includes the registers that have side-effects on the instruction execution as well as the particular fields of the register the instruction execution reads (is dependent on) and writes (updates). 

As \autoref{lst:mret-sailor} shows, Sailor also captures the footprint of the \texttt{sret} and \texttt{uret} instructions in addition to the footprint of exception handling on RISC-V. 
This is an artifact of the Sail model implementation. 
In the Sail model, there is a single \texttt{exception\_handler} function which is called by the instruction definitions of \texttt{mret}, \texttt{sret} and \texttt{uret}, as well as for other exception handling, with a switch case ladder that determines what code to execute in each of the case. 

\begin{lstlisting}[language=sailor-output,caption={ISA-insights captured by Sailor for \texttt{mret}.},captionpos=b,label={lst:mret-sailor}]
mret user executable: false
mret supervisor executable: false
mret machine executable: true 

mret ISA-state footprint: 
(includes mret execution path, 
interrupt handling path, and PMU) 
mstatus[MIE] R+W, mstatus[MPIE] R+W, 
mstatus[MPP] R+W, misa[U] Read,
mstatus[MPRV] Write, mie Read, 
mip Read, mideleg Read, mcountinhibit 
Read, instret Read, minstret Read, 
cycle Read, mcycle Read

(Sail implementation artifact: also 
includes sret, uret and exception
handling execution paths)
medeleg Read, mcause[IsInterrupt] Write, 
mcause[Cause] Write, mtval Write, mepc 
Write, mcause Read, mtvec Read, 
mstatus[SPIE] R+W, mstatus[UPIE] R+W, 
mstatus[SIE] R+W, mstatus[UIE] R+W, 
mstatus[SPP] R+W, sedeleg Read, 
misa[S] Read, misa[N] Read, stval 
Write, sepc Write, scause Read, 
utval Write, uepc Write, ucause Read, 
stvec Read, utvec Read, sideleg Read, 
misa[MXL] Read,	misa[C] Read, satp Read, 
mstatus[MXR] Read, mstatus[SUM] Read


\end{lstlisting}

The code block in \autoref{lst:mret} shows only the code for \texttt{mret} for simplicity, whereas in the model, the \texttt{exception\_handler} function is called in the else block of the code. 
In this particular case, using a single function does not help reducing redundant code, but rather reduces the modularity of the implementation. 
Further, the Scanner doesn't evaluate branches. 
So the Scanner captures a union of the footprint of all possible executions of an instruction. 
Thus, the Scanner over-estimates the ISA-state that can be influenced by an instruction. 
The benefit of the over-estimation is that the Scanner captures all the ISA-state that could be read/written by an instruction in any instance of the instruction's execution.
The drawback, which is a result of the way the Sail model is implemented, is that this over-estimation might include ISA-state that would never be read/written during any instance of the instruction execution. 
If there were different functions implemented for each \texttt{mret}, \texttt{sret} and \texttt{uret}, the footprint captured by the Scanner would be cleaner. 

Once we have the output from the Scanner, we provide it as input to the next component of the  tool, \ie{} the Analyzer.

\subsection{Analyzer} 
\label{sec:analyzer}

\begin{figure*}[t]
    \centering
    \includegraphics[width=350pt]{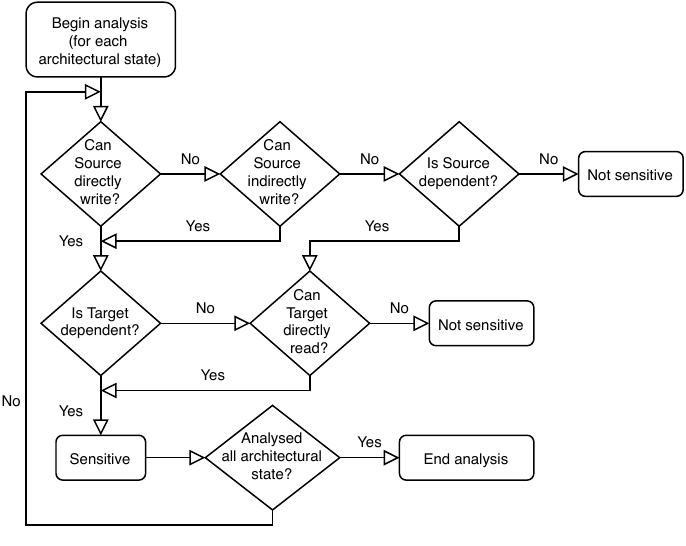}
    \caption{Algorithm to classify architectural state as security-sensitive. Source and Target correspond to privilege modes of the security domains involved in the context switch.}
    \label{fig:algo}
\end{figure*}

\autoref{fig:algo} shows the \textit{Classifier algorithm} we introduce to determine which ISA-state is security-sensitive. 
This analysis must be repeated for each pair of privilege modes in order to determine the exact ISA-state required to be swapped on a security domain switch. 
In \autoref{fig:algo}, the term \textit{Source} refers to the privilege mode in which the previous security domain had been executing, and \textit{Target} refers to the privilege mode in which the next security domain will be executing. 

The algorithm's goal is to identify any channels in the ISA-state that (i) allow the Source to affect the results of execution in the Target (adversarial Source), (ii) leak information of the Source to the Target (adversarial Target), or (iii) enable establishing communication (adversarial Source and Target). 
The analysis excludes GPRs since they can be written and read by any privilege mode and an adversary can easily leak data of another security domain through the GPRs (forming side-channels, covert-channels or computational integrity attacks), so they are considered security-sensitive by default. 
The analysis is repeated for each ISA-state (\eg{} a CSR). 

To identify (i), the algorithm checks whether the ISA-state is explicitly or implicitly writable by the Source mode. 
Here, explicit access refers to the first question in \autoref{sec:scanner-insights} (\eg{} \textit{csrw mstatus, in\_reg} instruction in RISC-V). 
Whereas, implicit access refers to the side-effects captured in the ISA-state footprint of an instruction (\eg{} the \texttt{fmadd} instruction in RISC-V execution can update the \texttt{mstatus.FS} bits to Dirty). 
If a privilege mode has the rights to execute any instruction that can write the ISA-state as a side-effect of execution, it is considered to have implicit write access to the ISA-state. 
The algorithm further checks whether the execution in Target mode depends on the ISA-state (via explicit or implicit reads). 
This ISA-state is considered as security-sensitive because if it is not appropriately swapped during the context switch, it can constitute \textbf{computational integrity} attacks on the Target security domain, breaching its integrity guarantees. 
For example, the Source can maliciously modify the rounding mode of computations, or disable extensions used by the Target.

Similarly, to identify (ii) and (iii), the algorithm checks whether the Source mode can implicitly or explicitly write the ISA-state and whether the Target mode can implicitly or explicitly read that ISA state. 
If the criteria is satisfied, then the ISA-state is considered as security sensitive because either the Target domain can breach the confidentiality of the Source domain by reading security-sensitive data of the Source domain or inferring details about execution in the Source domain. 
That would constitute \textbf{side-channel} attacks on the Source domain. 
Further, this ISA-state can also be used to breach the isolation guarantees in the system, enabling two-way communication between the Source and Target (\ie{} \textbf{covert-channel}). 

The algorithm performs one further check to identify (ii), \ie{} whether the ISA-state contains sensitive data of the Source that it depends on (implicitly reads) during execution (\eg{} \texttt{satp} CSR in RISC-V which holds the page table root address and is used implicitly during execution of load/store instructions).
If the Target can read this ISA-state, it can again constitute side-channel attacks on the Source domain.

If no such channels are identified, the algorithm considers the particular ISA-state as not security-sensitive and moves on to analyze another ISA-state. 

Despite the apparent simplicity of the questions that the algorithm asks to identify the potential information flow channels in the ISA-state, when combined with the fine-grained ISA-level insights that the Scanner generates, the Analyzer can help catch bugs that can constitute different classes of security vulnerabilities (such as computational integrity, side-channels, covert-channels).  

Once we have information regarding which ISA-state is security sensitive and which isn't, we can validate the existing context switch implementations against that information. 
We analyse the context switch implementations in existing security monitors to check whether they swap all the security-sensitive ISA-state.

We found three different classes of mishandled ISA-state across four different security monitors (Section \ref{sec:vuln-det}).

\subsection{Validator} 

\begin{figure}
    \includegraphics[width=\columnwidth]{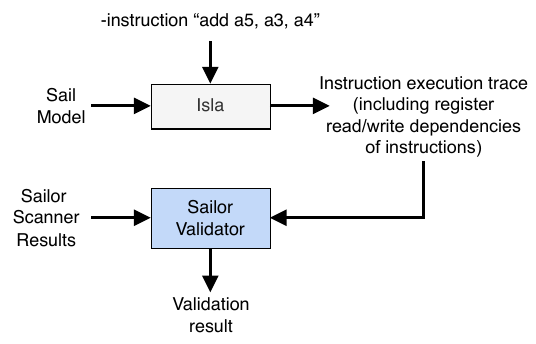}
    \caption{Sailor Validator}
    \label{fig:validator}
\end{figure}

\autoref{fig:validator} shows the third component of our tool, the Validator. 

The Validator's goal is to ensure that the Scanner captures all ISA-level insights from Sail using an existing tool Isla.
This is important as the goal of our tool is to help prevent any oversights on ISA-state in the context switch implementations.
Isla~\cite{isla} is a symbolic execution engine for Sail. It includes a footprint generator that produces instruction execution traces in the SMT-LIB2 format~\cite{smtlib2}.
Isla prunes exploration paths by assuming concrete values for the ISA-state.
Given an instruction and ISA-state constraints (e.g., privilege mode or CSR values), Isla generates traces capturing the instruction's dependencies, including the ISA-state that the instruction reads or writes.

\begin{lstlisting}[language=smtlib, caption={The SMT-LIB2 trace generated by Isla for the MRET instruction executed in M-mode.}, captionpos=b, label={lst:mret-isla}]
(trace
  (define-enum |Privilege| 3 (|User| |Supervisor| |Machine|))
  (assume-reg |cur_privilege| Machine |)
  (read-reg |PC| v442)
  (read-reg |cur_privilege| Machine |)
  (read-reg |mstatus| (struct (|bits| #x0000000a00000000)))
  (write-reg |mstatus| (struct (|bits| #x0000000a00000000)))
  (read-reg |mstatus| (struct (|bits| #x0000000a00000000)))
  (write-reg |mstatus| (struct (|bits| #x0000000a00000080)))
  (read-reg |mstatus| (struct (|bits| #x0000000a00000080)))
  (write-reg |cur_privilege| User |)
  (read-reg |mstatus| (struct (|bits| #x0000000a00000080)))
  (read-reg |misa| (struct (|bits| #x8000000000141101)))
  (write-reg |mstatus| (struct (|bits| #x0000000a00000080)))
  (read-reg |cur_privilege| User |)
  (read-reg |mstatus| (struct (|bits| #x0000000a00000080)))
  (write-reg |mstatus| (struct (|bits| #x0000000a00000080)))
  (read-reg |mepc| #x0000000000000000)
  (read-reg |misa| (struct (|bits| #x8000000000141101)))
  (write-reg |nextPC| #x0000000000000000)
  (define-enum |Retired| 2 (|SUCCESS| |FAIL|))
)
\end{lstlisting}

\autoref{lst:mret-isla} shows the trace generated by Isla for the execution of the \texttt{mret} instruction in M-mode. 
Along with the SMT-LIB2 semantics, there are additional annotations in the trace to describe the events meaningfully \eg{} \texttt{read-reg} or \texttt{write-reg} for a register read or write. 
We parse the trace and capture all the \texttt{read-reg} and \texttt{write-reg} events to generate the architectural footprint of the instruction. 
Then the Validator essentially checks the footprint generated from the Scanner against the one from Isla, as shown in Figure \ref{fig:validator}. 

The traces from Isla also capture if a trap occurs during the execution. 
For example,~\autoref{lst:mret-sup-isla} shows the execution of the \texttt{mret} instruction in S-mode, which is illegal according to the specification (~\autoref{lst:mret}). 
As shown in the trace, after reading that the \texttt{cur\_privilege} is S-mode (line 5 in the trace), the exception delegating register on RISC-V (\texttt{medeleg}) is read, the \texttt{mcause} register is written with the value 2 (line 9) which denotes an illegal instruction trap. 
More trap information is saved in the \texttt{mepc} and the \texttt{mtval} registers, and \texttt{cur\_privilege} is updated to M-mode so the trap can be handled in M-mode. 
In this way, we use Isla to also check which privilege modes are allowed to execute which instructions and directly read or write which CSRs. 
We compare the Scanner's output with Isla's for this information as well. 
Section \ref{isla-vs-sailor} discusses the comparison between Isla and Scanner.

\begin{lstlisting}[language=smtlib, caption={The SMT-LIB2 trace generated by Isla for the MRET instruction executed in S-mode.}, captionpos=b, label={lst:mret-sup-isla}]
(trace
  (define-enum |Privilege| 3 (|User| |Supervisor| |Machine|)) 
  (assume-reg |cur_privilege| Supervisor |)
  (read-reg |PC| v442)
  (read-reg |cur_privilege| Supervisor |)
  (read-reg |medeleg| (struct (|bits| #x0000000000000000)))
  (read-reg |misa| (struct (|bits| #x8000000000141101)))
  (read-reg |mcause| (struct (|bits| #x0000000000000000)))
  (write-reg |mcause| (struct (|bits| #x0000000000000002)))
  (read-reg |mstatus| (struct (|bits| #x0000000a00000000)))
  (read-reg |cur_privilege| Supervisor |)
  (read-reg |mstatus| (struct (|bits| #x0000000a00000000)))
  (write-reg |mstatus| (struct (|bits| #x0000000a00000800)))
  (write-reg |mtval| #x0000000000000000)
  (write-reg |mepc| v442)
  (write-reg |cur_privilege| Machine |)
  (read-reg |mcause| (struct (|bits| #x0000000000000002)))
  (read-reg |mtvec| (struct (|bits| #x0000000000000000)))
  (define-enum |TrapVectorMode| 3 (|TV_Direct| |TV_Vector| 
  |TV_Reserved|)) 
  (write-reg |nextPC| #x0000000000000000)
  (define-enum |Retired| 2 (|SUCCESS| |FAIL|))
)
\end{lstlisting}

\section{Implementation}

\subsection{Scanner and Analyzer Implementation}
We implement Sailor in Python. 
We use the Python lexical analysis library (plex) \cite{plex} to implement a parser for the Sail language, which is used by our Scanner.
Plex enables defining tokens, such as instruction definitions or CSR read/write expressions, using regular expressions, and associating each token with an action, like capturing CSR reads in an instruction's ISA-state footprint.
In Sail, each instruction has a definition, that further calls functions defined in the Sail model.
There are two approaches to the Scanner implementation: bottom-up (\ie{} find all the pieces of code in the Sail model where an ISA-state is modified and then trace back to find all the places where those pieces of code are called and so on) or top-down (scan one instruction definition at a time and track all the function calls it makes, while discovering the ISA-state being read/written). 
We implement the Scanner in a top-down fashion to limit recursion, focusing on computing the footprint of an instruction at a time, progressing with the natural execution flow. 

The Scanner first discovers all ISA-state definitions, function definitions and instruction definitions. 
Then it scans the function definitions and captures two things: the ISA-state footprint of the function (ISA-state read/written) and the function call footprint of the function (all the functions that are called inside the definition of this function).
Once all the functions have been scanned, the ISA-state footprint of each of the function is updated to add the ISA-state footprint of all the functions it calls. 
The function ISA-state footprints are cached to avoid redundant scanning. 
Then the Scanner parses all the instruction definitions, again capturing the ISA-state footprint and function call footprint for each instruction.
Similarly to functions, the ISA-state footprint of each instruction is updated to add the ISA-state footprint of all the functions it calls. 
The Scanner outputs the results in the CSV format.
The top-down approach and caching mechanism avoids recursion in our implementation of the Scanner. 
Python is also used to implement the Analyzer (\autoref{fig:algo}). 
\subsection{Validator Implementation}

As \autoref{fig:validator} shows, the Isla footprint tool generates the instruction execution trace per instruction. 
We automate the process of generating traces from Isla for many instructions.
We get the stream of instructions from the official instruction set listings in the RISC-V specification manual~\cite{riscv-specs} in the AsciiDoc format. 
We construct instructions from the listings which provides concrete values for the 7-bit opcodes and the associated operations (funct3 and funct7). 
For the source and destination registers, we use two concrete values from the GPRs in RISC-V: \texttt{zero} (x0) and any other GPR (\eg x10). 
We do this instead of using symbolic values here to limit the amount of path exploration in Isla. 
However, capturing the behaviour of the instruction with the read-only \texttt{zero} register in RISC-V ensures that the traces we capture covers this special case. 
For the immediate operands, we use symbolic values. 

We add a new option in Isla footprint to take as input the asciidoc file with the instruction listings and produce the traces for all the instructions. 
It does so in an annotated SMT-LIB2 format as \autoref{lst:mret-isla} shows. 
The Validator (written in Python) parses the instruction traces to extract the ISA-state footprint from Isla's output and compare those with the Scanner's output. 

\autoref{tab:loc} shows the code size of Sailor.\footnote{We plan to open-source Sailor.} 

\begin{table}[t]
\begin{center}
\begin{tabular}{ | c | c | } 
  \hline
  Module & Code Size (LoC)  \\ 
  \hline
  Scanner & 441 \\ 
  \hline
  Scanner RISC-V & 480 \\ 
  \hline 
  Analyzer & 181  \\ 
  \hline 
  Validator & 566 \\
  \hline 
\end{tabular}
\caption{Sailor code size in lines of Python code (LoC).
}
\label{tab:loc}
\end{center}
\end{table}

\subsection{Architecture Independent Implementation} 
\label{sec:arch-indep}
We apply Sailor primarily on the RISC-V Sail model \cite{sail-riscv}. 
However, Sailor can also be used with the Sail models of other architectures. 
Most of the Scanner implementation is architecture-independent. 
The Sail parser we implement using Plex is reusable across Sail models for other architectures such as Arm. 
We have an ISA-specific backend in the Scanner that is used to discover ISA-state \eg{} look for register definitions and find bitfields in the registers if they exist, etc. (see Scanner RISC-V in \autoref{tab:loc}). 
This is the only part that would need to be reimplemented to extend Sailor to other ISAs. 
The Analyzer implementation is completely architecture independent and can be used as is to analyze results from the Scanner for any ISA. 
The Validator implementation that compares that Scanner's output with that of Isla's traces is also architecture-independent. 
Isla can generate traces for Arm and RISC-V, so the validator can work on either of these.

\section{Evaluation}
\label{sec:eval}

The evaluation of Sailor answers the following questions: 

\begin{enumerate}
    \item Sailor's performance: Can Sailor scan the entire ISA specification in Sail and apply the algorithm to detect security-sensitive ISA state in a reasonable time? 
    \item Vulnerability detection: How can Sailor Analyzer's results be applied to existing or new security-critical systems for detecting architectural vulnerabilities? 
    \item Validating the Scanner: Does the Scanner correctly capture all ISA-level insights from Sail? 
\end{enumerate}

\myparagraph{Evaluation setup:} To answer above questions, we ran Sailor on Ubuntu 22.04.5 LTS with Linux v5.15 on a computer equipped with with an AMD EPYC 7642 processor with 48 cores, 256\,GB RAM, and 2\,TB AVAGO MR9461-16i hard drive. We ran Isla on MacOS Monterey v12.7.5
on a computer equipped with a 2.9 GHz Quad-Core Intel Core i7, and 16\,GB RAM.

\begin{figure*}
    \includegraphics[width=\textwidth]{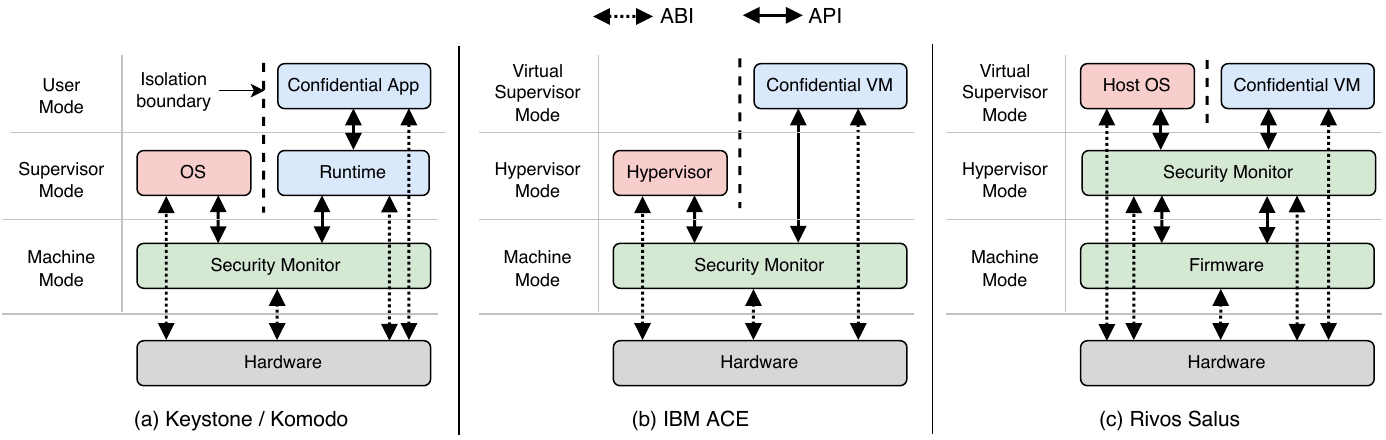}
    \caption{Overview of the confidential computing frameworks.}
    \label{fig:cc-frameworks}
\end{figure*}

\subsection{Sailor's performance}
The Scanner parses the RISC-V ISA model in Sail, which comprises of 23.5 kLoC (kilo lines of code) and writes the collected insights into CSV files. 
The Analyzer uses those CSV files as input to execute the algorithm shown in \autoref{fig:algo} and determine for each architectural state whether it is security sensitive. 
The analysis in Sail was performed on the base ISA \ie{} rv64gc, along with the following extensions: vector, user-interrupts, scalar cryptography ISA extensions. 
The Scanner parsed 355 instruction definitions and 160 CSRs (including bitfields derived from 62 actual CSRs) from the RISC-V Sail model~\cite{sail-riscv}.  
For a secure context switch from S-mode to U-mode, 70 of the CSRs/bitfields are classified as security sensitive by the Analyzer. 
We also applied Sailor on the Sail implementation of the Hypervisor extension~\footnote{This extension is in the process of being upstreamed, but is functional enough to boot up Linux-based system on the generated emulator.}~\cite{sail-hext-git, sail-hext}. 
The complete execution of the Scanner and the Analyzer takes 6 seconds in both cases. 

\subsection{Vulnerability detection} 
\label{sec:vuln-det}
We used the results from the Analyzer 
to corroborate the context switch implementations in existing systems: Keystone~\cite{keystone}, Komodo~\cite{komodo}, Rivos' Salus~\cite{rivos-salus}, and IBM's ACE~\cite{ibm-ace}, see \autoref{fig:cc-frameworks}. 
We found three classes of mishandled ISA-state that when exploited, leads to five different types of security vulnerabilities. 
\autoref{tab:vuln} summarizes these findings. 
We manually analyzed the context switch code in all the frameworks. 
We also generated automated tests for Keystone to check whether the security-sensitive ISA-state before and after a context switch is the same.

\begin{table*}[t]
\begin{center}
\begin{tabular}{ | c | c | c | c |} 
  \hline
  Mishandled ISA-state & Framework & Context-switch action & Security vulnerability\\ 
  \hline
  \multirow{2}{*}{\texttt{senvcfg} CSR} & Keystone & \multirow{2}{*}{Does not swap} & \multirow{2}{*}{Misconfigured emulated I/O fences} \\ 
  \cline{2-2} 
   & Komodo & & \\
  \hline 
  \multirow{2}{*}{F-extension} & Keystone & Does not swap & Computational integrity*, Side-channel*, Covert-channel \\ 
  \cline{2-4}
  & Rivos Salus & Swaps conditionally & Timing-channel \\
  \hline 
  V-extension & Rivos Salus & Swaps conditionally & Timing-channel \\
  \hline 
\end{tabular}
\caption{Mishandled ISA-state found using Sailor and the corresponding security vulnerabilities it constitutes. (*) These two security vulnerabilities were previously discovered using manual analysis of the ISA \cite{dtrap-fpu}.}
\label{tab:vuln}
\end{center}
\end{table*}

\subsubsection{Keystone}
Keystone \cite{keystone} is a security monitor that enables creating trusted execution environments (TEEs) on the RISC-V platform. 
Keystone enclaves comprise of a trusted runtime that executes in S-mode and an enclave application that executes in U-mode. An untrusted OS executes in S-mode as usual. 

We analyzed the security domain switch between the enclave and the untrusted OS as a switch from S-mode to S-mode via the security monitor in M-mode. 
So we query Sailor to produce security sensitive ISA state with both source and target privilege modes as S-mode. 
We then use those insights to verify that the context switch code in Keystone correctly saves all security sensitive state for the source security domain and similarly restores all security sensitive state for the target security domain. 

Our analysis found several security-sensitive ISA-state that Keystone does not swap during the context switch. 

We found that Keystone \textbf{does not swap the \texttt{senvcfg} CSR}. 
The \texttt{senvcfg} CSR has a \texttt{FIOM} (Fence of I/O implies fence of Memory) bitfield. 
The \texttt{FIOM} bit exists to handle emulated I/O for a VM running in U-mode and a hypervisor running in S-mode when paravirtualization is not available. 
When such a VM has multiple virtual harts, concurrent accesses to the device by the harts translate to concurrent accesses to memory due to the device emulation. 
However, since the VM is not aware of the device emulation, it might simply execute I/O fences to order accesses to the device. 
When the \texttt{FIOM} bit is set, these I/O fence instructions lead to memory accesses also being ordered via memory fences. 
If this bit is used by the trusted runtime in Keystone to emulate a device for the enclave application, any transitions to the untrusted OS can modify and reset this bit. 
If the trusted runtime transitions to the OS for unrelated reasons, it is unlikely for the runtime to enforce checks on the architectural state that it expects the security monitor to take care of saving and restoring during the context switch. 
Further, the OS can maliciously slow down an enclave that doesn't use the \texttt{FIOM} feature, by enabling it and requiring any I/O fences that execute inside the enclave application to be paired with memory fences. 

Further, the Floating Point extension (F-extension) in RISC-V is included in a general-purpose core (rv64gc) by default. 
However, Keystone does not swap the ISA state corresponding to the F-extension, introducing leaks across security domains. 
This includes the \texttt{fcsr} control and status register (superset of the \texttt{fflags} and \texttt{frm} CSRs) and general-purpose floating point registers \ie{} \texttt{f0}-\texttt{f31}.  
This opens up the attack surface due to the following potential security vulnerabilities.

The \texttt{fcsr} register has a floating point rounding mode field (\texttt{frm}).
\textbf{Computational integrity} attacks can be mounted by the untrusted OS by maliciously modifying the \texttt{frm} field (floating point rounding mode) in the \texttt{fcsr}, that leads to varying computation results inside the enclave. 
Similarly, the general-purpose floating point registers \ie{} \texttt{f0}-\texttt{f31} can also be leveraged to breach the computational integrity of the enclave.

\textbf{Side-channel} and \textbf{covert-channel} attacks can also exploit the F-extension CSRs in RISC-V.
These attacks can leverage the 
\texttt{fcsr} register, specifically the \texttt{fflags} field, which records floating-point exceptions like divide-by-zero. Instead of raising traps, the processor updates \texttt{fflags}, requiring software to proactively check for exceptions. 
These attacks can also leverage the \texttt{frm} field and the \texttt{f0}-\texttt{f31} registers.

Prior work has already discovered two of the vulnerabilities we describe above (the side-channel and computational integrity attacks through the \texttt{fcsr} register \cite{dtrap-fpu}) and reported them to Keystone \cite{keystone-fpu-issue}. 
Although Keystone is currently under active development, the authors have neither responded to the report nor introduced a fix in the framework. \footnote{We reproduce the side-channel and computational integrity attacks by prior work \cite{dtrap-fpu} and also provide a proof-of-concept implementation for new side-channel and covert-channel attacks that leverage the \texttt{f0}-\texttt{f31} registers.}

\subsubsection{Komodo} 
Komodo is a verified security monitor to create enclaves~\cite{komodo}. 
Komodo (initially developed on Arm) has been ported to RISC-V and verified by a framework called Serval~\cite{serval}. 
It has a model similar to Keystone, with a security domain switch occurring from S-mode to S-mode. 

The Serval framework verifies Komodo for the RISC-V integer, multiplication, and Zicsr extensions. 
Considering only these extensions, we still found that Komodo does not swap the \textbf{\texttt{senvcfg} CSR}. 

However, we note that the \texttt{senvcfg} CSR was introduced into the RISC-V privileged specification v1.12, which was ratified after the publication of the paper~\cite{serval, riscv-specs}. 
But if Komodo is deployed on more recent platforms, an adversary can exploit the bug.

This insight shows that it is important to automate detection of security-sensitive ISA-state as it is a process that must keep up with the constantly changing ISA specifications. 

\subsubsection{Rivos' Salus}
Rivos Salus is a security monitor that executes in the HS-mode on RISC-V and creates confidential VMs that execute in VS-mode. 
In this case, we analyze the security domain switch between the confidential VM and the untrusted host OS as a switch from VS-mode to VS-mode through the Rivos security monitor in HS-mode. 

We found that Salus correctly swaps all security-sensitive ISA state correctly in the context switch implementation. 
However, it introduces a potential \textbf{timing channel} vulnerability by not swapping all security-sensitive ISA state during all instances of context switch. 
For both, vector and floating point extensions, Salus swaps the architectural state only if the state is dirty. 
This can leak information about the use of extensions inside the confidential VM to the untrusted OS due to varying delay in the context switch. 
Even though traditionally it is recommended to only switch registers pertaining to such extensions~\cite{context-switch-cost-hotos} if they were used, as a performance optimization, this can hamper security in confidential computing frameworks. 
Avoiding that requires constant-time context switching, albeit at the cost of performance.

\subsubsection{IBM's ACE}
IBM's ACE~\cite{ibm-ace} is a framework to create confidential VMs on RISC-V. It implements the RISC-V CoVE specification~\cite{riscv2025cove} and its security monitor runs in M-mode, the untrusted hypervisor in HS-mode and confidential VMs in VS-mode, as \autoref{fig:cc-frameworks} shows. 

We found that IBM's ACE correctly swaps all security-sensitive ISA state correctly in the context switch implementation. 
However, we found that ACE also swaps ISA-state that is not security sensitive for a switch between HS-mode and VS-mode. 
For instance, M-mode specific CSRs that are nor accessible to either HS-mode or VS-mode, neither influence execution in these privilege modes. 
Examples include: \texttt{mtval} and \texttt{mtval2} (that contain exception-specific information). 
Further, HS-mode CSRs that do not influence execution in VS-mode and are not accessible in VS-mode also do not require to be swapped, specifically when switching from HS-mode to VS-mode.
Examples include \texttt{htval} and \texttt{htinst}.
Since security domain context switches are on the critical path and expensive \cite{context-switch-cost, context-switch-cost-hotos}, swapping only security-sensitive ISA-state can help reduce the cost. 
We recommend optimizing context switch performance by reducing the amount of non security-sensitive ISA-state that is swapped during context switches, over only swapping security-sensitive ISA-extension state when it is in use. 
\footnote{Since ACE is under active development and has not been tested on real hardware yet, we consider the performance analysis out of scope for this paper.}

\subsubsection{ISA extensions}
\label{sec:isa-ext-discussion}
While Keystone doesn't claim support for platforms with extensions such as vector (V), user interrupt and exception handling (N), it takes no preventative measures to detect an unsupported platform and halt further execution or disable those extensions. 
Similarly, IBM's ACE, Rivos Salus and Komodo, also do not take any measures to prevent being incorrectly deployed on unsupported platforms. 
The task of deploying a system only on suitable platforms is left for the system operators/platform providers to take care of. 
Adding such checks can add defense-in-depth and proactively prevent exploitation of any security vulnerabilities that stem from architectural extensions not supported by security-critical software like security monitors, especially for platforms like RISC-V that encourage custom extensions. 

Further, it can be beneficial w.r.t. security to adhere to the following practices while designing new ISA extensions.
ISA-state previously not classified as security-sensitive must not be affected by the extension to be reclassified as security-sensitive. If that is the case, then Sailor needs to be reapplied on the entire ISA specification to correctly determine security-sensitive ISA-state. 
Newly introduced ISA-state should not be security-sensitive when the extension is disabled. 
This is important to ensure that privileged software that does not support certain extensions can still correctly execute on the platform by disabling those extensions.

\subsection{Validating the Scanner} 
\label{isla-vs-sailor}

\myparagraph{Isla and Scanner performance comparison}: We generated Isla traces for the rv64gc ISA. To collect the traces that generate the same amount of insights as the Scanner, Isla takes 53 minutes. Even though the performance evaluation of Scanner has been carried out on a more powerful machine, it is unlikely that generating traces from Isla can be sped up to the same magnitude as generating insights from Sailor (6 seconds). 
Using the Scanner is beneficial in real-time, (\eg{} in CI/CD pipelines of security monitors under active development) especially if the software evolves with the evolving ISA specifications. 
Whereas, we can still leverage Isla to corroborate the results of the Scanner offline, which provides an additional confidence in the correctness of the tool. 

\myparagraph{Isla and Scanner insights comparison}: Comparing the output of the Scanner with the output from the Isla traces, the Validator identified one bug in the Scanner. 
A function call was not added appropriately to the interrupt analysis, ultimately failing to capture a write in the \texttt{mstatus.MIP} bit in the trace of three instructions (store word (\texttt{sw}), store double word (\texttt{sd}), and store conditional (\texttt{sc})). 
Apart from this bug, we successfully validated the remaining results that the Scanner generates against Isla's traces. 

The Scanner's output must be verified to be a superset of Isla's output. 
This is because the Scanner captures a union of the footprint of all possible executions of an instruction. 
Wheread Isla provides the trace for a single instance of the execution.
For example, \autoref{lst:mret-isla} and \autoref{lst:mret-sup-isla} are two instances of execution of the same instruction, but with different privileges. 
The trace in \autoref{lst:mret-sup-isla} reports a write to \texttt{mcause}, while the trace in \autoref{lst:mret-isla} does not. 
The Scanner captures the write to \texttt{mcause} regardless.

Further, the Sail model implementation sometimes groups similar instruction into a single definition (\eg{} \texttt{fadd}, \texttt{fsub}, \texttt{fdiv}, and \texttt{fmul}) and then branches out into different functions. 
In these cases, we first compute a union of the footprints we capture from the Isla traces for the group of instructions before comparison.

\section{Related Work}

Prior work includes manual analysis of the ISA specification to find exploitable bugs in context switch implementations~\cite{guard-dilemma, rop-sgx, dtrap-fpu, totw, enclave-isolation}. 
Manual analysis of the entire ISA specification can trivially become erroneous, similar to introducing bugs in context switch implementations in the first place. 
These prior works typically focus on particular systems to analyze or finding bugs related to particular ISA extensions (\eg{} Floating Point Unit~\cite{dtrap-fpu}) to limit the scope of the required manual ISA-analysis. 

Some privileged software are formally verified for functionality and security properties, such as non-interference, often requiring multiple years of effort for extensive verification~\cite{sel4, certikos, komodo}. 
However, even that is not always enough to guarantee strict isolation at the ABI layer as these works write their own system specifications which are usually a subset of the actual ISA. 
This is done in order to make the verification feasible, since the work tends to be strenuous and if the state explored is not limited, then the verifiers or provers can quickly face path explosions. 
For instance, prior work verifies security monitor software for RISC-V and x86-32 architectures, but model only the integer (I), multiplication (M) and control and status register (Zicsr) instruction sets for RISC-V, and only GPRs for x86-32~\cite{serval}. 

Sail has an Ocaml parser that can be used with any of the different backends also written in Ocaml. 
The backends are for compiling Sail into the C/Ocaml emulators, or producing theorem prover definitions for Coq, Hol4, Isabelle, etc. 
In the existing implementation, the Sail parser cannot directly produce the type of ISA insights we extract with the Scanner. 
There are existing efforts to create a new backend that works with the Ocaml parser to extract ISA-related information from Sail \cite{thinkopenly}. 
The type of ISA-related information extracted includes name and operands of instructions, their description and register names, descriptions and sizes. 
The backend is meant to be simple and serve as a stepping stone for researchers to be able to extend that into more sophisticated use-cases. 
While Sailor is focused towards extracting more specific information from Sail towards the pre-defined goal of secure context switch implementations, compared to the work described above, both the works are aligned in making use of Sail to automatically retrieve ISA-related information and replace manual efforts of browsing through ISA specifications. 

Prior work involves attacks at the API-layer (as \autoref{fig:conf-comp} shows) which includes exploiting missing pointer checks, buffer overflows, etc. that can cause memory leaks or breach integrity of security domains~\cite{iago, totw}. 
Our work specifically focuses on the ABI-layer and not the API-layer, and thus requires mitigation strategies to prevent attacks at the API-layer to work in tandem. 
Similarly, our work does not target preventing micro-architectural attacks (cache side-channels, speculative execution attacks~\cite{spectre, meltdown, flush-reload}) and requires separately applying mitigation strategies for those as well.

\section{Future Scope}

As we describe in \autoref{sec:arch-indep}, Sailor is architecture-indepedent apart from an ISA-specific backend for the Scanner. 
While RISC-V serves as a suitable candidate for demonstration, the challenge of manually analyzing complex ISAs like Arm and x86 is significantly greater. Sailor can significantly expedite the process of identifying security-sensitive ISA-state for those ISAs. 
Extending Sailor to support other ISAs remains a future work.  

The Analyzer's output \ie{} security-sensitive ISA-state, can be leveraged by verification frameworks such as Serval ~\cite{serval} to ensure that they prove security properties on the context switch code with an ISA model including all the security-sensitive ISA-state. 
The Analyzer's output can also be used to exclude the ISA-state that is not security-sensitive from the verification frameworks's ISA specification model. 

Sailor can be integrated with the CI/CD pipelines of security monitors or other privileged software to enforce checks on the context switch implementation against Sailor's results. 
This can be even more useful for software under active development as Sailor can proactively capture insights from the latest ISA specifications or as the software starts supporting more ISA extensions.

\section{Conclusion}

Our work concentrated on an often overlooked area of the attack surface: the ABI-layer. 
The increasing complexity of ISA specifications make it more difficult to write secure context switch code. We introduced a tool, Sailor, that leverages machine-readable ISA specifications written in the Sail language. 
Our tool parses Sail code to identify which ISA-state must be swapped during a context switch across different security domains. 
Consequently, our tool replaces the strenuous task of manually navigating through hundreds or thousands of pages of ISA specifications to figure out which ISA-state to swap during the context switch. 
We detected three classes of mishandled ISA-state across four confidential computing frameworks for the RISC-V ISA using the results from our tool. 
Two of the three classes we identify stem from ISA-extensions. 

As ISAs continue to evolve, it is crucial to automate the task of analysing the latest specifications. 
We believe that as ISAs continue to evolve and increase in complexity, tools like Sailor are crucial for automating tedious manual tasks. 
Further, it is important to keep the machine-readable ISA specifications in Sail up to date as well, rather than just relying on prose-style documentations. 
We encourage hardware vendors to provide Sail models for their machines, especially when the hardware implements custom ISA-extensions, to promote comprehensive security analysis of the hardware.

\section*{Acknowledgments}
We thank Silvio Dragone (IBM Research Zurich), Edouard Bugnion (EPFL, Switzerland), Michael Le, Elaine Palmer (IBM T. J. Watson Research Center), Lennard Gäher (MPI-SWS Germany), as well as Charly Castes and Adrien Ghosn (EPFL, Switzerland) for their valuable comments and feedback.
This work was conducted as part of the project on Assured Confidential Computing (ACE) for RISC-V.

\bibliographystyle{plain}
\bibliography{refs}

\end{document}